\documentclass[conference]{IEEEtran}
\IEEEoverridecommandlockouts
\pdfoutput=1
\usepackage{cite}
\usepackage{amsmath,amssymb,amsfonts}

\usepackage[noend]{algpseudocode}
\usepackage{algorithmicx,algorithm}

\usepackage{graphicx}
\usepackage{subfigure}
\usepackage{textcomp}
\usepackage[hyphens]{url}
\usepackage[hyphenbreaks]{breakurl}
\usepackage{algorithm,algpseudocode}
\usepackage{xcolor}
\def\BibTeX{{\rm B\kern-.05em{\sc i\kern-.025em b}\kern-.08em
    T\kern-.1667em\lower.7ex\hbox{E}\kern-.125emX}}
\begin{document}

\title{An Open Case-based Reasoning Framework for Personalized On-board Driving Assistance in Risk Scenarios\\
}

\author{\IEEEauthorblockN{Wenbin Gan, Minh-Son Dao, Koji Zettsu}
\IEEEauthorblockA{\textit{Big Data Integration Research Center} \\
\textit{National Institute of Information and Communications Technology (NICT)}\\
Tokyo, Japan \\
$\{wenbingan,dao,zettsu\}@nict.go.jp$}

}

\maketitle

\begin{abstract}
Driver reaction is of vital importance in risk
scenarios. Drivers can take correct evasive maneuver at proper cushion time to avoid the potential traffic crashes, but this reaction process is highly experience-dependent and requires various levels of driving skills. To improve driving safety and avoid the traffic accidents, it is necessary to provide all road drivers with on-board driving assistance. 
This study explores the plausibility of case-based reasoning (CBR) as the inference paradigm underlying the choice of personalized crash evasive maneuvers and the cushion time, by leveraging the wealthy of human driving experience from the steady stream of traffic cases, which have been rarely explored in previous studies. To this end, in this paper, we propose an open evolving framework for generating personalized on-board driving assistance. In particular, we present the FFMTE model with high performance to model the traffic events and build the case database; A tailored CBR-based method is then proposed to retrieve, reuse and revise the existing cases to generate the assistance. We take the 100-Car Naturalistic
Driving Study dataset as an example to build and test our framework; the experiments show reasonable results, providing the drivers with
valuable evasive information to avoid the potential crashes in different scenarios.

\end{abstract}


\section{Introduction}

\begin{figure}[t]
\centerline{\includegraphics[width=0.4\textwidth]{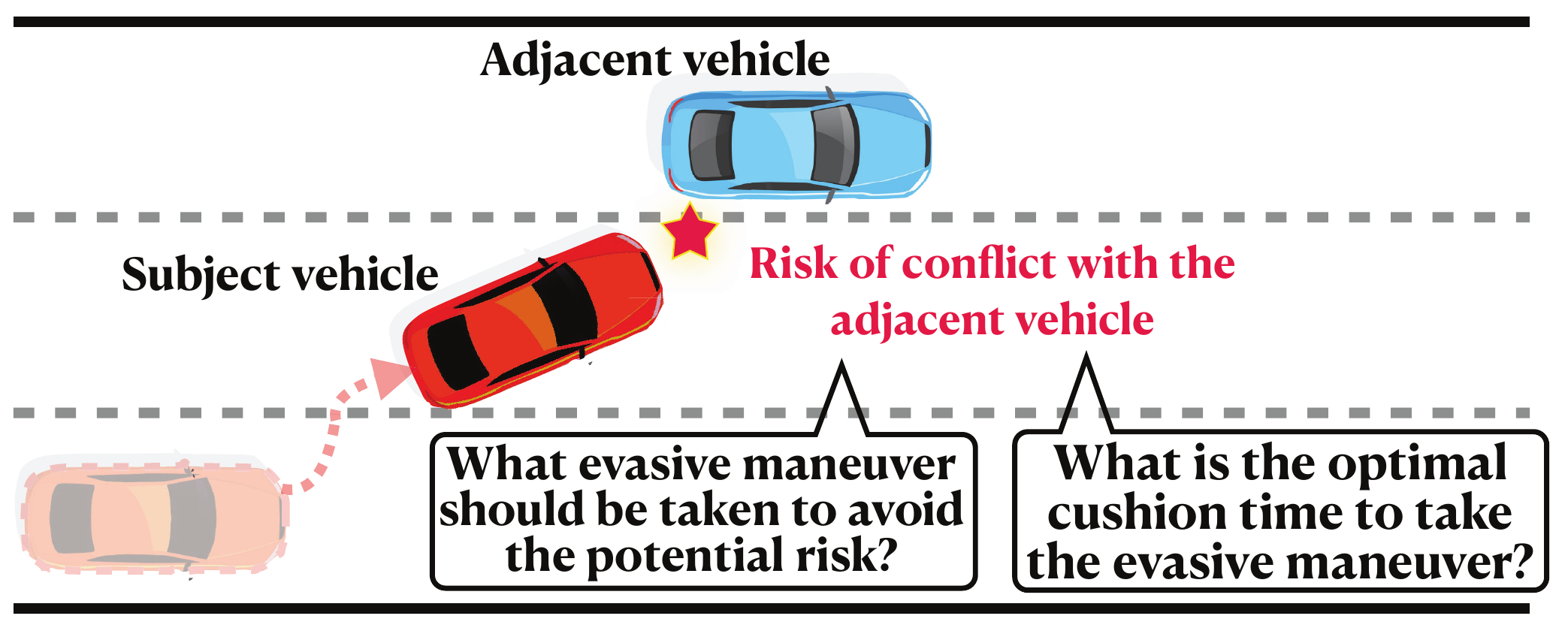}}
 \vspace{-5 pt}
\caption{An example of a risk scenario: the subject vehicle changes lane and triggers the risk of potential conflict with an adjacent vehicle.}
\label{Research_question}
\end{figure}

Reducing traffic risks is one of the most important public safety concerns across the world. The global statistics by World Health Organization show that more than 1.35 million humans lost their lives every year due to the road traffic accidents \cite{world2019global}. Contrary to other calamities, it is well acknowledged that the traffic crashes are not chance events \cite{ahmed2022global} and hence the majority of traffic crashes can be mitigated (avoided) via appropriate driving maneuvers\footnote{A maneuver is denoted as a driver's reaction in response to the precipitating event in a risk scenario. Braking, decelerating, accelerating and steering are common reactions.} at proper operation time 
\cite{wang2020safedrive,khan2019net}. As shown in Figure \ref{Research_question}, the subject vehicle (SV) changes lane and triggers the risk of potential conflict with an adjacent vehicle; if the SV steers to the right side at a proper short cushion time after detecting this risk scenario, the risk could be avoided. For this proactive crash mitigation purpose, some advanced driving assistance systems (DASs) that offer safety recommendations have been proposed in this driver-vehicle co-driving situation, and showed significant safety improvement \cite{wang2019crash,yi2020implicit,ahmed2022global}. 

DASs have been greatly boosted by recent developments in sensing and computing technologies, and dynamic mobile communication between vehicles (V2V), vehicles and road infrastructures (V2I) \cite{perumal2021insight,abdelrahman2022robust}. These systems assist drivers in plenty of aspects to perceive the contexts and make proper decisions thereby minimizing the driving risks and saving human lives, money and time \cite{yi2020implicit,perumal2021insight}. Given these substantial benefits,  various researches have been conducted to predict the driving risks \cite{abdelrahman2022robust,chen2019driving,wang2020safedrive}, analyze driving behaviors \cite{sama2020extracting,bouhoute2018advanced}, recommend speed and steering control
\cite{saito2021context,koh2015integrated}, and assist lane keeping and changing \cite{yi2020implicit},  etc. These researches have great potential to improve the driver vigilance so as to avoid the potential risk of crashes. In fact, completely avoiding the traffic risk is still unrealistic at this point \cite{wang2019crash}. Several studies in the US show that driver error is the critical reasons for appropriate 94\% of crashes, with 33\% attributing to human decision error \cite{singh2018critical,perumal2021insight}. 
Therefore, determining what evasive maneuver should be performed to hinder the crash occurrence or at least minimize the crash severity, and also what is the optimal cushion time to take the corresponding maneuver, are of great importance in these inevitable risk scenarios. 
This is also the research questions that we want to solve in this paper. 

Driving is a highly experience-dependent activity that requires various levels of  driving skills, physical and cognitive characteristics of human drivers \cite{sama2020extracting, horswill2021thousand}. 
The experience can be mutually imparted among human drivers and has been verified to effectively improve the on-road driving safety after learning \cite{horswill2021thousand}.  
When facing a new traffic scenario, human drivers tend to rely on their experience to understand and perform similar operations to make reactions, especially in the short-term, sudden risk scenarios \cite{sama2020extracting,kolodner1992introduction}. Hence for the inexperienced drivers, the on-board driving assistance in these risk scenarios appears to be particularly important. Actually, many researches have reported that novice drivers between the age of 15-29 have the highest numbers of crash events \cite{world2019global,chen2019driving}, and inexperienced drivers have a higher crash risk than their counterparts \cite{evans2020young}. 
Moreover, professional drivers may occasionally make mistakes while attempting to prevent a crash \cite{perumal2021insight}. Given the severe loss of human lives and money, the better leverage of human driving experience to provide driving assistance to all road drivers is promising to lower the crash rate considerably in these risk scenarios. 

Meanwhile, recent technological developments in intelligent vehicles and transportation offer new wealthy sources of data about the drivers, the vehicles and the context information, making the building of a holistic view of natural driving roadway events (under the risk scenarios) both technologically possible and economically feasible. These large numbers of event cases contain valuable driving experience from numerous drivers on their operations to hinder the crash occurrence or reduce crash severity, which can be potentially used as human experience to instruct the drivers in the on-board DASs. However, these cases have rarely been used in existing researches to infer the optimal evasive maneuver and the cushion time in the risk scenarios. 

This paper presents an open CBR-based framework for
on-board driving assistance in risk scenarios, with the aim of inferring the optimal crash avoidance maneuver and cushion time to hinder the crash occurrence. We assume that the risk scenarios have already been detected using the LiDAR sensing or computer vision techniques \cite{wang2020safedrive,chen2019driving}. 
We take the ``100-Car Naturalistic Driving Study (NDS)'' dataset as an example to build our framework, which consists of three steps: the traffic event modeling, the case database building, the CBR-based on-board driving assistance. We present a field-aware factorization machine (FFM) \cite{juan2016field} based approach, namely FFMTE, to model the traffic events considering the event information, driver-vehicle interaction, road condition and the driving context, and build the overall case database. Personal driving records are also stored into a personal driving case database to provide further individualized adaption. We propose a maneuver inference method based on the case-based deductive reasoning to reuse the  existing cases to solve the current risk scenario. The personalized on-board driving assistance is generated by revising the similar cases to adapt to the individual drivers based on the driving context and their own driving history. 

The contributions of this work are threefold:
\begin{itemize}
\item We propose an open evolving framework for on-board driving assistance by leveraging the wealthy of human driving experience from the steady stream of traffic cases, which have been rarely explored in previous studies.
\item We present the FFMTE model with high performance to model the traffic events and build the overall case database. The dense embeddings of variables learnt though FFM also facilitate the case retrieval in the subsequent CBR process.  
\item We propose a tailored CBR-based method to reuse and revise the similar cases to generate the personalized on-board
driving assistance.
\end{itemize}

\section{Related Work}

\subsection{Crash Mitigation and Avoidance Technologies}

Accident mitigation and avoidance
systems are becoming increasingly common in the commercial vehicles. 
These systems enhance the driving safety by warning drivers to the potential risks or automatic take actions to avoid or mitigate the crashes \cite{yi2020implicit,mahmoudzadeh2019studying}. 
Khan et al. \cite{khan2019net} analyzed the effectiveness of three crash avoidance technologies (blind-spot monitoring, lane departure warning, and forward-collision warning) in the U.S. for the year 2015, and 
concluded that the three technologies could collectively prevent up to 1.6 million crashes each year including 7200 fatal crashes, which validated the great potential of such systems to save human lives, time and money.  

Active and passive strategies are both applied for the crash mitigation and  avoidance \cite{wang2015driving}. 
Passive strategy is primarily concerned with improving automobile safety features such as seat belts, airbags, and body structures to mitigate the severity in the inevitable crashes.
Active methods proactively detect and evaluate the driving risks \cite{wang2020safedrive,wang2015driving,bouhoute2018advanced,chen2019driving}, recommend or take actions to control the vehicles to avoid and mitigate crashes \cite{wang2019crash,koh2015integrated,huang2022efficient,sama2020extracting}. Among them, driving risk prediction and analysis is the hottest topic and initial step in this direction because only the risk scenarios are detected can the further actions be taken to avoid the potential risks. Wang et al. \cite{wang2020safedrive} proposed a deep learning model to handle both spatial and temporal dependence in driving risk evaluation incorporating the driving patterns of the target vehicle, the driving patterns of surrounding vehicles, and the interactions between the target and surrounding vehicles to improve driving safety. Chen et al. \cite{chen2019driving} presented a deep autoencoder based method for driving safety risk prediction. By adaptively searching the optimal sliding window size and extracting the features on the original driving data, their method showed good prediction performance on the driving risk. 

After detecting the potential risks, various action-taking methods are proposed to avoid and mitigate the crashes.  
Wang et al. \cite{wang2019crash} designed a 
motion planner for generating an emergency path to mitigate the inevitable crash as much as possible. 
Morales et al. \cite{morales2017proactive} proposed a model for velocity control by abstracting the maneuvers from expert driving data to avoid the potential crashes in the blind interactions. 
Wang et al.  \cite{wang2018learning} provided a learning-based method to understand the driving style and offer the warning to avoid the lane departure crashes. 
Kaplan and Prato \cite{kaplan2012application} applied the regret minimization as behavioral paradigm to choose the crash avoidance maneuvers.

Some researches have studied the correlations between the performing of crash avoidance maneuvers and various factors, providing the evidences to choosing proper maneuvers under multiple precrash conditions.  
Wang et al. \cite{wang2016correlation}
examined the correlations between various crash avoidance maneuvers and injury severity sustained by motorcyclists. 
Ahmadreza et al. \cite{mahmoudzadeh2019studying} discovered the effect of distraction-related factors on the performance of a crash avoidance avoidance maneuver, paving the way for help the drivers to react better. 

This paper focuses on the action-taking
method to avoid and mitigate the crashes, a step further the risk situation being  detected. Different from the existing studies, we propose an open evolving framework for crash avoidance by leveraging the wealthy of human driving experience from the steady stream of traffic cases, which is the first work in this direction to use the overall human knowledge for crash avoidance.

\begin{figure*}[th]
	\centering
	\subfigure[distribution of two categories]{
		   \includegraphics[width=0.29\textwidth]{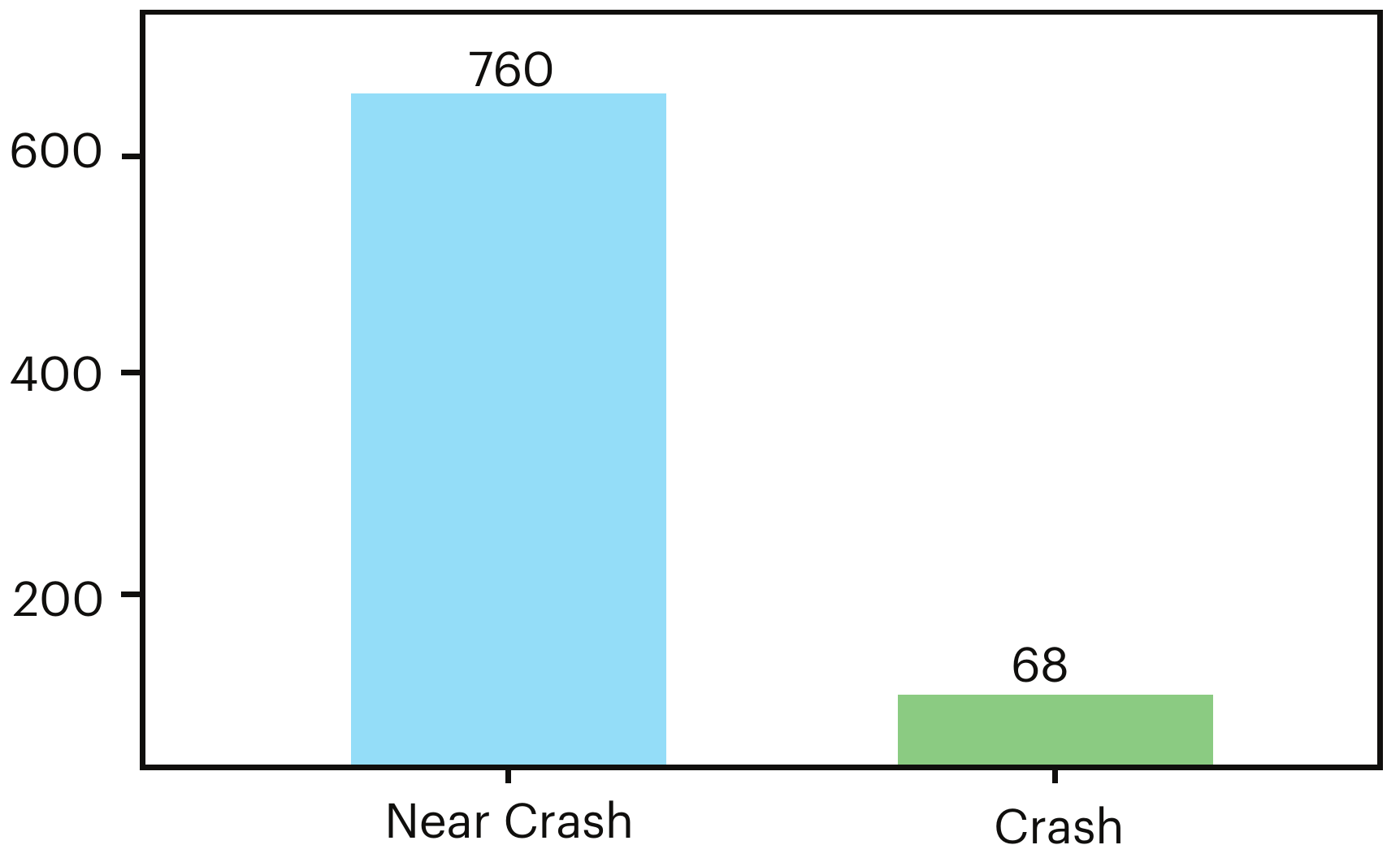}
		\label{fig:distribution_data}
	}
	\subfigure[distribution of two kinds of events]{
   		 	\includegraphics[width=0.3\textwidth]{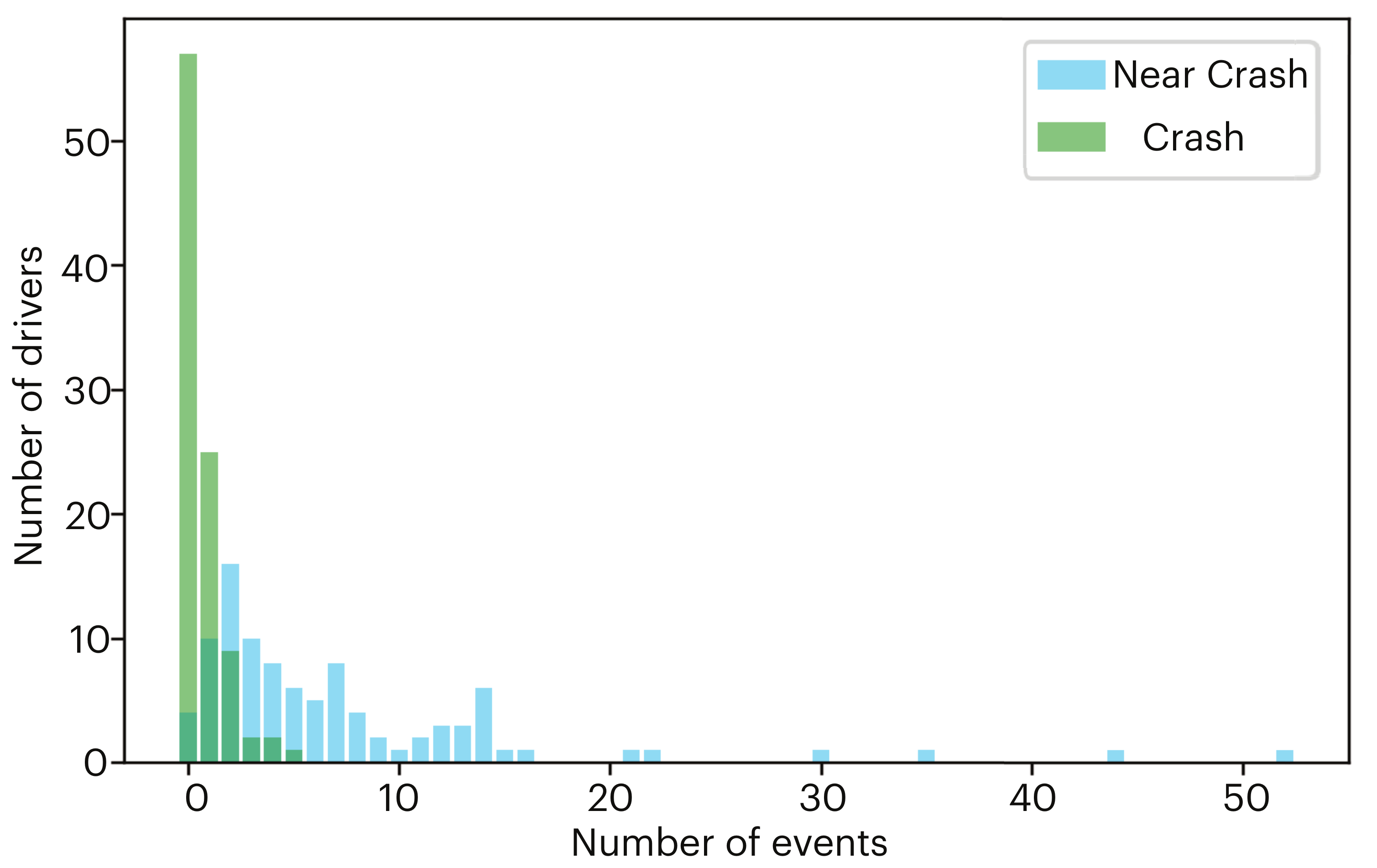}
		\label{fig:hist_graph_event}
    }
    \subfigure[distribution of cushion time]{
   		 	\includegraphics[width=0.35\textwidth]{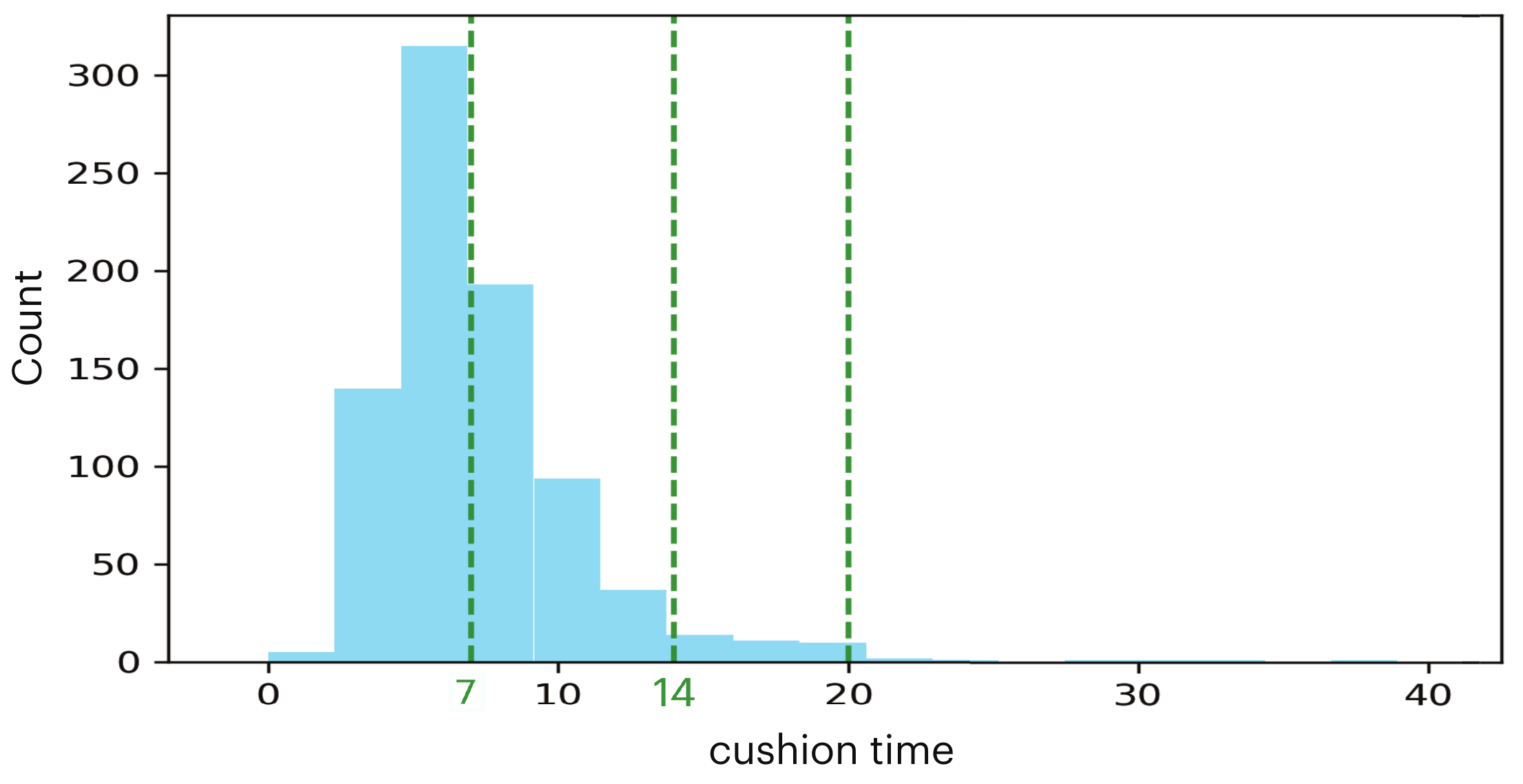}
		\label{fig:cushion_time}
    }
     \vspace{-5 pt}
	\caption{The distributions of the statistical information for the 100-Car NDS dataset}
	\label{fig:dataset}
\end{figure*}

\subsection{Driving Assistance Systems}
DASs support drivers by relaxing their control efforts, improve their risk perception, warn them in the case of a mistake, provide required assistance, and are proved to be potential for safe and comfortable driving \cite{khan2019comprehensive}. 
Depending on their functionalities, the current DASs can be divided into three categories: the safe driving support systems (SDS), the driver monitoring systems (DMS), and the in-vehicle information systems (IVIS) \cite{yi2020implicit}.
SDSs control the vehicle dynamics, proactively forecast the driving risk,  reduce and avoid the potential crashes. Typical systems of this kind are blind-spot monitoring \cite{morales2017proactive}, lane departure warning \cite{wang2018learning}, and collision avoidance \cite{wang2019crash,kaplan2012application}, cruise control \cite{koh2015integrated}, lane changing and interaction assistance \cite{perumal2021insight}. 
DMSs monitor the driver behavior and driving status to provide the warning. For example, 
Horberry et al. \cite{horberry2021human} presented a multi-modal driver fatigue and distraction warning (visual, auditory and tactile) to help reduce impairment-related incidents. Driving behavior, style and affective states are also studied in some researches \cite{sama2020extracting,abdelrahman2022robust}. IVISs provide the in-vehicle  services and the interested information to  drivers, such as the weather and traffic information, the route planning and entertainment services. 

This research is a part of the SDSs, with a special focus on choosing the crash avoidance maneuvers and the optimal operation time in the risk scenarios for promoting the safer driving.

\subsection{CBR for Intelligent Vehicles}

Case-based reasoning uses old experiences similar to the current one to understand and solve new problems \cite{kolodner1992introduction}.
For its effectiveness in deductive reasoning, it has been used in various knowledge-based systems \cite{louati2021deep,bouhana2013integrated,kolodner1992introduction}. 

For intelligent vehicle applications, some researches have explored its potential from various aspects. 
Louati et al. \cite{louati2021deep} used CBR for controlling the traffic signal to efficiently guide emergency vehicles through intersections and avoid the disturbances of the traffic flow.
Bouhana et al. \cite{bouhana2013integrated} integrated CBR with Choquet integral to suggest the itinerary that best matches the user's preferences when facing a new situation. 
Olivier et al. \cite{quirion2021case} adopted CBR to design new routes to fulfill orders by retrieving and adapting routes previously performed by the on-line retailers to effectively address the last-mile delivery problem. 
Ma et al. \cite{ma2014automated} automated the generation of traffic incident response plan based on CBR and Bayesian theory to guide responders to take actions effectively and timely after traffic incidents.

These existing studies have explored the potential of using the CBR-based methods to control the traffic signal and plan route, etc. However, few have investigated the CBR for crash mitigation and avoidance, a gap bridged in this paper.

\section{Data Description and Event Case Abstraction}
This paper takes the 100-Car Naturalistic Driving Study (NDS) dataset\footnote{The 100-Car NDS data: \url{https://dataverse.vtti.vt.edu/dataset.xhtml?persistentId=doi:10.15787/VTT1/CEU6RB}} as an example to build our evolving framework. 

\subsection{Dataset}
The 100-Car NDS \cite{neale2005overview} is the first large-scale instrumented vehicle study conducted in the Northern Virginia/ Washington, D.C. area over a 2 year period with the primary purpose of collecting large-scale, naturalistic driving data. 
Naturalistic driving data depicting drivers' daily driving in urban regions were unobtrusively collected from vehicles devised with 
a suite of sensors including forward and rearward radar, lateral and longitudinal accelerometers, gyro, GPS, the vehicle CAN, and five channels of video cameras.
The dataset includes approximately 2,000,000 vehicle miles, almost 43,000 hours of data, 102 primary drivers, 12 to 13 months of data collection for each vehicle. 

From the data, an event database was created with three types of safety-related events  identified: crashes, near-crashes, and other safety-critical incidents \cite{neale2005overview}.  
A crash is defined as ``any collision between the subject vehicle and another item.'' A near-crash is  ``a conflict scenario that necessitates a quick, severe evasive maneuver to prevent a crash.'' The quick, evasive maneuver includes actions using steering, braking, accelerating, or any combination of control inputs \cite{chen2019driving}.
A safety-critical incident is an incident of lesser magnitude than a near crash. The naturalistic approach makes it possible to record all kinds of these events. 

\begin{table*}[t]
\caption{Definition of the proposed data  transcription protocol.}
 \vspace{-5 pt}
\label{transcription_protocol}
\resizebox{\textwidth}{!}{ %
\begin{tabular}{lll} \hline
\textbf{Variable}                               & \multicolumn{1}{l}{\textbf{Meaning}}                     & \textbf{Description}\\ \hline
{\textbf{Event Information}}      &&\\
\qquad Event Severity (2){*}                                            & \begin{tabular}[c]{@{}l@{}}valid triggered occurrences of \\ a near-crash or crash\end{tabular}
& 0: near crash, 1: crash\\
\qquad Event Nature (8)                                               & \begin{tabular}[c]{@{}l@{}}Specifies the type of potential crash \\ or near-crash that may occur\end{tabular}                                              & \begin{tabular}[c]{@{}l@{}}0: Conflict with a lead vehicle, 1: Conflict with vehicle in adjacent lane/Conflict with merging vehicle, \\ 2: Single vehicle conflict, 3: Conflict with a following vehicle, \\ 4: Conflict with vehicle moving/turning across/into another vehicle path (opposite/same direction), \\ 5: Conflict with obstacle/object (parked vehicle,pedalcyclist, animal,pedestrian) in roadway\\ 6: Conflict with oncoming traffic, 7: Unknown conflict,Other\end{tabular}\\
\qquad Precipitating Event (10)                                        & \begin{tabular}[c]{@{}l@{}}the critical event that triggered the \\ crash or near-crash possible.\end{tabular}                                             & \begin{tabular}[c]{@{}l@{}}0: Other vehicle ahead --stopped on roadway/decelerating/ slowed and stopped/at a slower constant speed,\\ 1: Subject ahead -- slowed and stopped 2 seconds or less/decelerating/stopped on roadway more than 2 seconds,\\ 2: Subject lane change \\ 3: Other vehicle lane change \\ 4: This Vehicle Lost Control -- Poor road conditions/Excessive speed/Other cause,\\ 5: Other vehicle entering intersection -- straight across path/left turn across path/turning same(opposite) direction,\\ 6: Subject in intersection -- turning right/ turning left/passing through,\\ 7: Subject over right/left edge of road, Subject over right/left lane line\\ 8: Object(Animal,Pedestrian,Pedalcyclist,etc.) in roadway, Animal in/approaching roadway\\ 9: Others\end{tabular} \\ \hline
{\textbf{Driver-Vehicle Interaction}} &&\\
\qquad Pre-Incident Maneuver (10)                                      & \begin{tabular}[c]{@{}l@{}}the last action that the subject vehicle driver \\ engaged in just prior to the precipitating event\end{tabular}                & \begin{tabular}[c]{@{}l@{}}0: Decelerating in traffic lane, 1: Going straight, accelerating, 2: Going straight, constant speed,\\ 3: Merging, 4: Starting in traffic lane, 5: Changing lanes, 6: Stopped in traffic lane,\\ 7: Negotiating a curve, 8: Turning right, 9: Turning left, \\ 10: Others(Making U-turn,Entering/Leaving a parked position,Backing up,etc.)\end{tabular}\\
\qquad Driver Reaction (7)                                            & \begin{tabular}[c]{@{}l@{}}The subject driver's reaction or evasive \\ maneuver in response to the precipitating event.\end{tabular}                       & \begin{tabular}[c]{@{}l@{}}0: Braking (no lockup,lockup,lockup unknown), 1: Braked and steered right/left, 2: Steered to left/right, \\ 3: Accelerated, 4: Accelerated and steered right/left, 5: No reaction, 6: Other actions\end{tabular}\\
\qquad Driving Cushion Time (4)                                       & \begin{tabular}[c]{@{}l@{}}the duration time from the beginning of precipitating \\ event until the completion of the final evasive maneuver.\end{tabular} & 0: 0$\sim$7s, 1: 7$\sim$14s, 2: 14$\sim$20s, 3: more than 20s\\ \hline
{ \textbf{Static Road Condition}}&&\\
\qquad Surface Conditions (5)                                         & The type of roadway surface condition& 0: Dry, 1: Wet, 2: Icy, 3: Snowy, 4: Others\\
\qquad Traffic Flow (3)                                               & Roadway design& 0: Divided (median strip or barrier), 1: Not divided, 2: Others (One-way traffic,No lanes, etc.)\\
\qquad Travel Lanes(8)                                               & The number of travel lanes& 0: 1, 1: 2, 2: 3, 3: 4, 4: 5, 5: 5, 6: 6, 7: Others\\
\qquad Traffic Control (3)                                            & Type of traffic control applicable to the vehicle& \begin{tabular}[c]{@{}l@{}}0: No traffic control, 1: Traffic signal\\ 2:Others (Traffic lanes marked,Stop sig,Yield sign,Officer or watchman,One-way road or street, etc.)\end{tabular}\\
\qquad Relation to Junction (2)                                       & Subject driver's relation to junction& 0: Non-Junction, 1: others (Intersection, Intersection-related,Entrance/exit ramp, Interchange Area, etc.)\\
\qquad Alignment (3)                                                  & \begin{tabular}[c]{@{}l@{}}Geographical description of the roadway \\ that best suits the condition\end{tabular}                                           & 0: Straight level, 1: Curve level, 2: Others (Straight grade,Curve grade, Straight hillcrest,etc.)\\
\qquad Locality (6)                                                   & Best description of the surroundings& \begin{tabular}[c]{@{}l@{}}0: Business/industrial, 1: Interstate, 2: Open Country, 3: Residential, \\ 4: Construction Zone, 5: Others(School,Church, etc.)\end{tabular}\\ \hline
{ \textbf{Dynamic Driving Context}}&&\\
\qquad Visual Obstructions (2)                                        & \begin{tabular}[c]{@{}l@{}}Visual factors that may have contributed to \\ the cause of the precipitating event\end{tabular}                                & \begin{tabular}[c]{@{}l@{}}0:No obstruction                                     \\ 1:Others (Sunlight glare, Parked vehicle, Building, Curve or hill,Trees, crops, vegetation, etc.)\end{tabular}\\
\qquad Traffic Density (6)                                            & \begin{tabular}[c]{@{}l@{}}The level of traffic density at the time of the \\ start of the precipitating event.\end{tabular}                               & \begin{tabular}[c]{@{}l@{}}0: Level-of-service A, 1: Level-of-service B, 2: Level-of-service C, 3: Level-of-service D\\ 4: Level-of-service E, 5: Level-of-service F\end{tabular}\\
\qquad Lighting (5)                                                   & Lighting condition& 0: Daylight, 1: Darkness, lighted, 2: Dusk, 3: Darkness, not lighted, 4: Dawn\\
\qquad Weather (6)                                                    & Weather condition& 0: Clear, 1: Cloudy, 2: Raining, 3: Mist, 4: Snowing, 5: Others\\
\qquad Number of Other Vehicles (5)                                   & \begin{tabular}[c]{@{}l@{}}Other than the subject vehicle, number of objects \\ involved in the crash or near-crash\end{tabular}                           & 0: 0, 1: 1,  2: 2, 3: 3, 4: 4\\ \hline
\multicolumn{3}{l}{*The number in the brackets is the total kinds of potential values for the variable.}\\

\end{tabular}}
\end{table*}

In this paper, we use the crashes and near-crashes data to explore the potential for crash mitigation and avoidance.  
All the 68 crashes and 760 near-crashes in the 100-Car NDS dataset are used in our framework. The collected event videos allow direct viewing of all of the pre-event and during-event parameters, based on which a event summary is created for each event  including the pre-event maneuver, precipitating factor, event type, contributing factors, associative factors and
avoidance maneuver, etc \cite{dingus2006100}. The detailed information on crashes and near-crashes demonstrates drivers' unsuccessfully/successfully performing an evasive maneuver or decision making to choose the incorrect/correct evasive maneuver, providing insight into developing and refining crash avoidance countermeasures for these risk scenarios \cite{dingus2006100}.

The distributions of the statistical information for the dataset are shown in Figure \ref{fig:distribution_data}, \ref{fig:hist_graph_event} and \ref{fig:cushion_time}. As shown in Figure 
\ref{fig:hist_graph}, most drivers did not have the crash event, 26\% drivers had one time of crash, and only one driver have 5 times of crash. In contract, most of the drivers had numerous near-crash events ranging from 1 to 16, with 52 the maximum. 
Figure \ref{fig:cushion_time} shows the distribution of the cushion time. Here cushion time is the time period from the beginning of the precipitating event to the completion of the final evasive maneuver, it indicates the constrained time required for the drivers to take the quick evasive maneuver to avoid a potential risk. It is calculated as follows:
\begin{equation}
    t_{cushion}= (f_{end}-f_{start})/r
\end{equation}
where $f_{start}$ and $f_{end}$ are the starting and ending points in the collected driving video when the sequences defining the occurrence of the event begins or ends. The $r$ is the frame rate of the video (i.e., 7.5 frames per second). As shown in Figure \ref{fig:cushion_time}, in most of the events, drivers took the evasive maneuver at 1$\sim$9 seconds.

\subsection{Abstraction of Driving Event Cases\label{sec:Abstraction_Event_Data}}

The original dataset provides a total of 57 variables to comprehensively describe an driving event. In this paper, we select parts of the significant variables that both satisfy our aim of inferring the proper evasive maneuver and the cushion time for crash mitigation and avoidance,
and also have been verified to contribute to the driving events in previous researches \cite{wang2015driving,abdelrahman2022robust,wang2020safedrive}. 

\begin{figure}[!t]
	\centering
	\subfigure{
		   \includegraphics[width=0.22\textwidth]{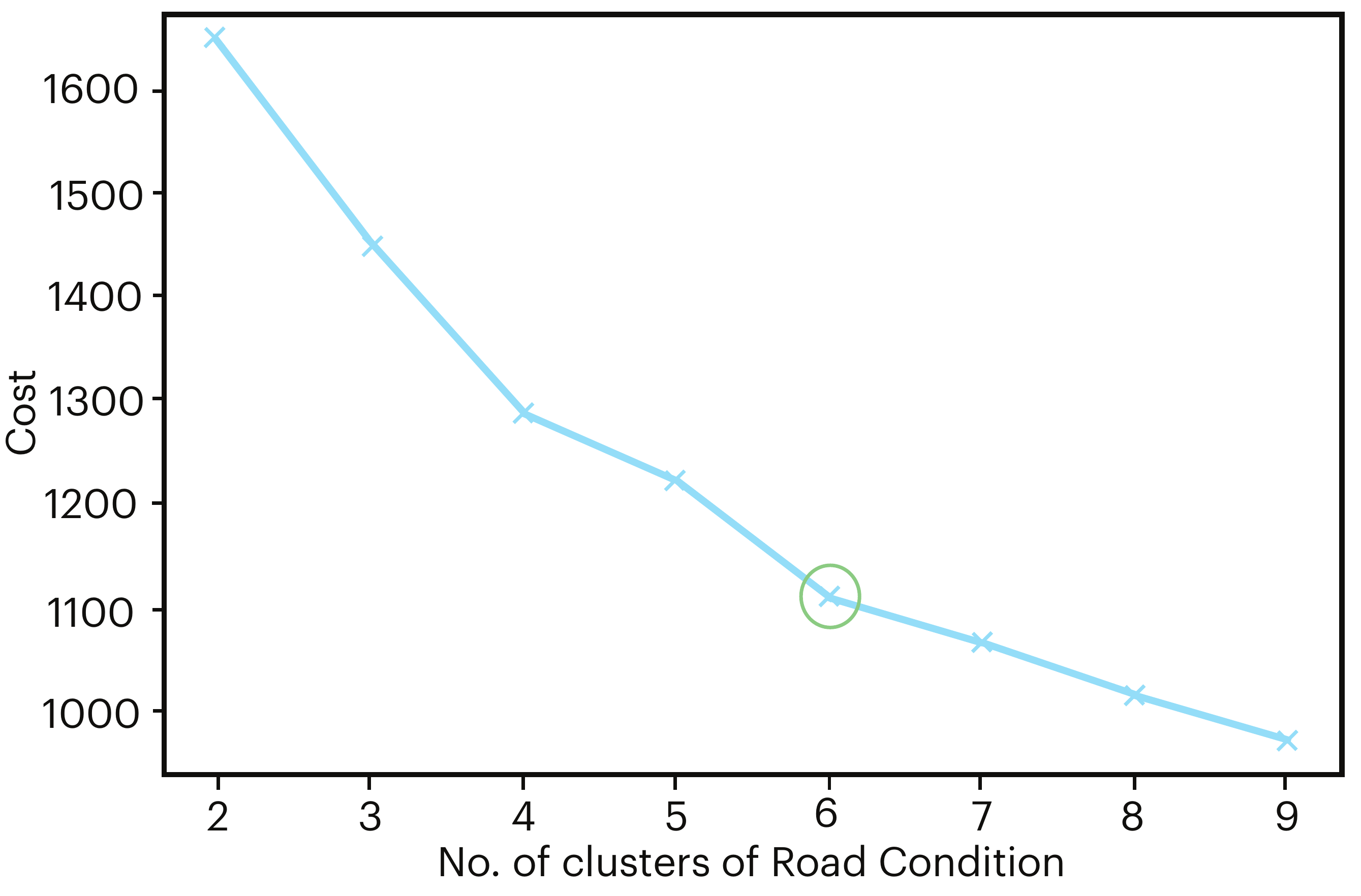}
		\label{fig:distribution}
	}
	\subfigure{
   		 	\includegraphics[width=0.22\textwidth]{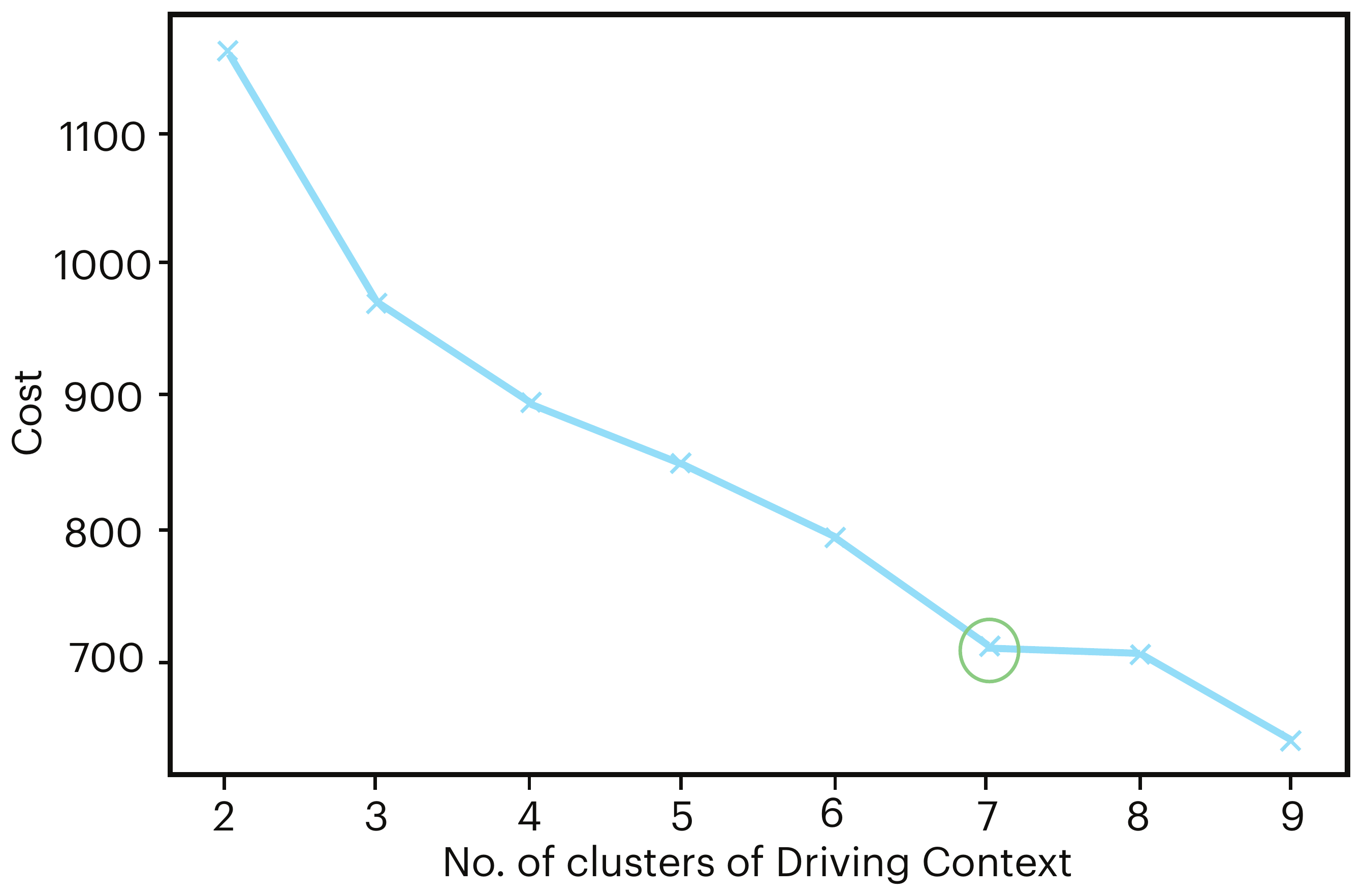}
		\label{fig:hist_graph}
    }
    \vspace{-5 pt}
	\caption{Clustering cost of road condition (left) and driving context (right) with different numbers of clusters.}
	\label{fig:Clustering}
\end{figure}

A novel data-transcription protocol that takes into account a broad range of significant variables that contribute to the driving events, is proposed in Table \ref{transcription_protocol}. 
It is divided into four primary categories: the event information (event severity, event nature, precipitating event), driver-vehicle interaction (pre-incident maneuver, driver reaction, driving cushion time), static road condition (surface condition, traffic flow, etc.), and dynamic driving context (visual obstructions, traffic density, etc.)
This protocol offers the capability for studying and explaining the relationship between driving events, driver/vehicle information, and static and dynamic surroundings.

As shown in Table \ref{transcription_protocol}, 
each of these variables contains several possible values, for example, the ``event nature'' contains eight values coding from 0 to 7. For the driving cushion time, as shown in Figure \ref{fig:cushion_time}, we abstract it into several intervals based on its distribution. Four intervals are obtained from the data: 0$\sim$7s, 7$\sim$14s, 14$\sim$20s, and more than 20s. 

To improve the model interpretability and reduce the further computing complexity for overall variables, we use k-modes \cite{huang1998extensions} to cluster the static road condition and dynamic driving context to further abstract the driving data.
The k-modes algorithm\footnote{The Python implementation of the k-modes algorithm for clustering categorical data is here: \url{https://pypi.org/project/kmodes/}} deals with category data using a simple matching dissimilarity measure, substitutes the means of clusters with modes, and updates modes in the clustering process using a frequency-based technique to minimize the clustering cost function \cite{huang1998extensions}. 
Basically, k-modes takes a dataset $X = (X_{1}, X_{2},..., X_{N}$) of $N$ points as input, each point is a multi-dimensional vector (7 for the static road condition and 5 for the dynamic driving context). It finally generates $k$ cluster centroids $C$ and assign each data point to a distinct cluster, as expressed in the follow formula:
\begin{equation}
    Minimize\sum_{i=1}^{k}\sum_{j=1}^{N}||X_{j}-C_{i}||
\end{equation}

Figure \ref{fig:Clustering} shows the results of k-modes clustering cost of road condition and driving context with different numbers of clusters. 
We choose $k=6$ for the road condition and $k=7$
for the driving context with an elbow method for the optimal clusters. 

\begin{figure}[!t]
\centerline{\includegraphics[width=0.35\textwidth]{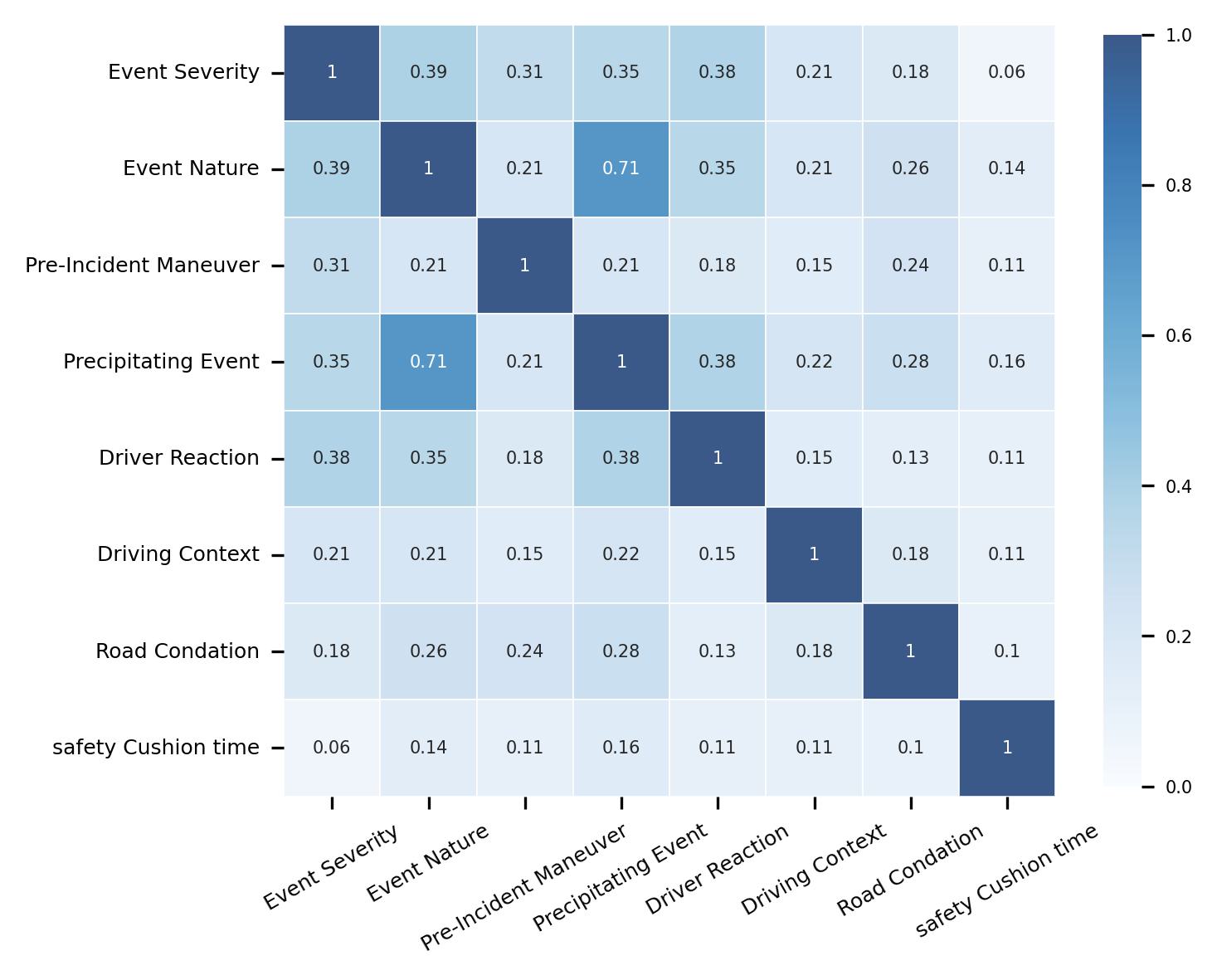}}
 \vspace{-5 pt}
\caption{Cramer's V correlation coefficient matrix.}
\label{cramers_V_correlation}
\end{figure}

Finally, each driving event is represented by eight variables in the four categories: event severity (e\_s, 2), event nature (e\_n, 8), precipitating event (p\_e, 10), pre-incident maneuver (p\_m, 10),  driver reaction (d\_r, 7), driving cushion time (c\_t, 4), static road condition (r\_c, 6), dynamic driving context (d\_c, 7). The expressions and numbers in the brackets are the abbreviations and the total kinds of potential values of the corresponding variables. Note that the variable ``driver reaction'' shows the evasive maneuver the driver taken to avoid a crash. 

\textbf{Definition 1 (Event Case Abstraction):} a event case is represented as $c = \{e\_n, p\_e, p\_m, d\_r, c\_t, r\_c, d\_c \}$ with the $e\_s$ as a label to indicate the case severity.

\begin{figure*}[!tbp]
\centerline{\includegraphics[width=\textwidth]{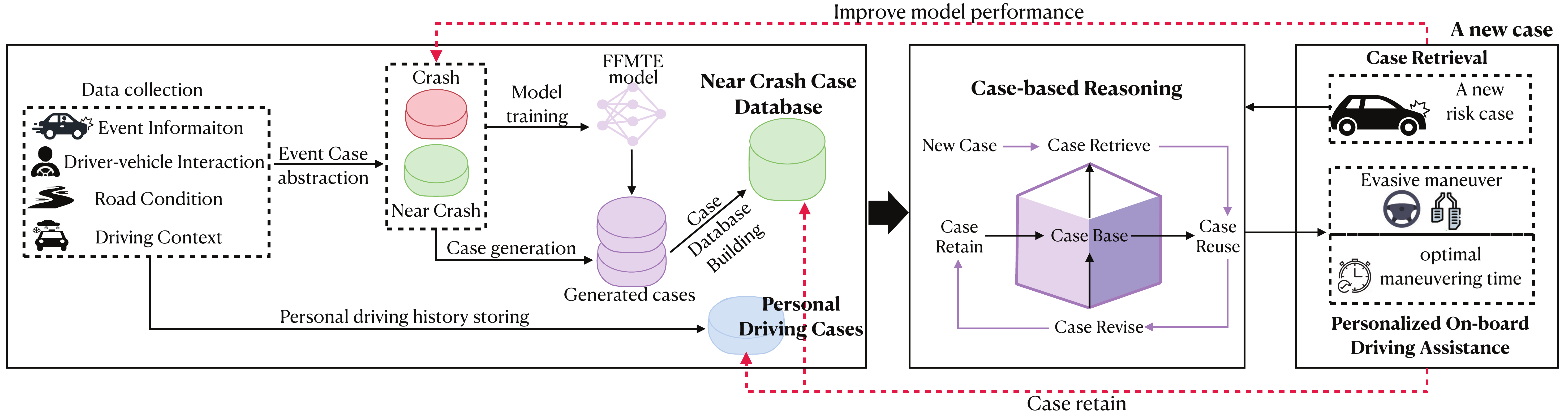}}
\vspace{-10 pt}
\caption{The proposed CBR-based evolving framework for the on-board personalized driving assistance in risk scenarios.}
\label{framework}
\end{figure*}

As all the eight variables are transformed into category ones, we use Cramer's V correlation to calculate the coefficient matrix, as shown in Figure \ref{cramers_V_correlation}. The event severity is correlated with all the other variables to different degrees, and these variables are also correlated with each other. For example, the precipitating events are highly correlated with the event nature, and also the road condition (icy, snowy) can also influence the driving context (traffic density). 
To better model the traffic events, such interactions among different variables must be considered.

\section{Our Proposed Framework}

Figure \ref{framework} shows the overall framework for the on-board personalized driving assistance in risk scenarios. This framework mainly includes three steps: 
the traffic event modeling, the case database building, the CBR-based on-board personalized driving assistance, combining which we show how to leverage the wealthy sources of human
driving experience from the steady stream of traffic cases to assist drivers to mitigate and avoid the potential crashes in the risk scenarios. 

For modeling the traffic event, we propose a the FFMTE model to model the severity (crash or near-crash) of each event. 
FFMTE takes consideration of multiple variables and also their interactions to classify the driving event, and also generates the dense embeddings for the considered variables, which can be used for the similarity calculation in the case retrieval. For the case database building, our idea is that the near-crash cases involves the valuable information about the evasive maneuvers and the cushion time that can be used to instruct future driving in the risk scenarios, hence we generate all the possible cases and use the trained FFMTE model to obtain all the near-crash cases to build the overall case database. Moreover, the personal driving event cases are also stored into the personal database. For the on-board personalized driving assistance, we propose a CBR-based method to retrieve and reuse the existing cases from the case database, and then use the personal driving history and the driving context to revise the reused cases to adapt to the individual situation and generate the personalized driving assistance. 
We claim this framework is an open one as it can evolve with increasingly driving event cases accumulated into the databases, and the naturalistic daily driving provides the steady stream of countless traffic cases thus making the framework more and more precise, a novelty of this work compared with all the existing studies.  
For reference, the algorithm for the proposed framework is shown in Alg. \ref{CBR_algorithm}.

\begin{algorithm}[tb]\small
\caption{The proposed CBR-based framework for on-board drving assistance}
\label{CBR_algorithm}
\begin{algorithmic}[1]
\Require All drivers' driving event data $X$, the personal driving event data $X_{i}$ for a driver in the risk scenarios, a new risk case $p_{i}$ for the driver. 
\Ensure The overall case base $CB$, the personal driving case base $CB_{i}$ for the driver, the evasive maneuver $d\_r_{i}$ and the optimal maneuvering time $c\_t_{i}$ for the risk case $p_{i}$. \\
\textbf{Traffic Event Modeling:}
\State label and abstract the event data $X$;
\State train the model $FFMTE \leftarrow X$;\\
\textbf{Case base building:}
\State generate all the possible cases $X^{'}$ using SMOTEN method: 
$X^{'} \leftarrow SMOTEN(X)$;
\State obtain all the near crash cases using the trained model  \{$X^{'}_{near-crash}$\}$\leftarrow$ $FFMTE(X^{'})$ and store them into the case base $CB$;
\State store the personal data $X_{i}$ into the personal case base $CB_{i}$;\\
\textbf{CBR for on-board personalized driving assistance:}
\State $\{d\_r_{i},c\_t_{i}\}$ $\leftarrow$ CBR($p_{i}, CB, CB_{i}$);\\
\textbf{Update the framework:}
\State  store the confirmed (crash/near-crash) new case $\{p_{i},d\_r_{i},c\_t_{i}\}$ to $X$, $CB$ and $CB_{i}$;
\State retrain the $FFMTE$ model for more accurate traffic event modeling;
\State repeat the above process to evolve the framework constantly;
\State return $\{d\_r_{i},c\_t_{i}\}$, $CB$, $CB_{i}$.
\end{algorithmic}
\end{algorithm}

\subsection{FFMTE Model for Traffic Event Modeling\label{sec:FFM_TE}}

\textbf{Formulation of Traffic Event Modeling:} Given an event dataset of pairs of $(c, e\_s)$, we want to find a function $f$ that can best model the relationship between the vector of variable predictors, $c$, and the severity of the driving event, $e\_s$: 
\begin{equation}
    f: c \rightarrow e\_s
\end{equation}

\begin{figure}[!tbp]
\centerline{\includegraphics[width=0.45\textwidth]{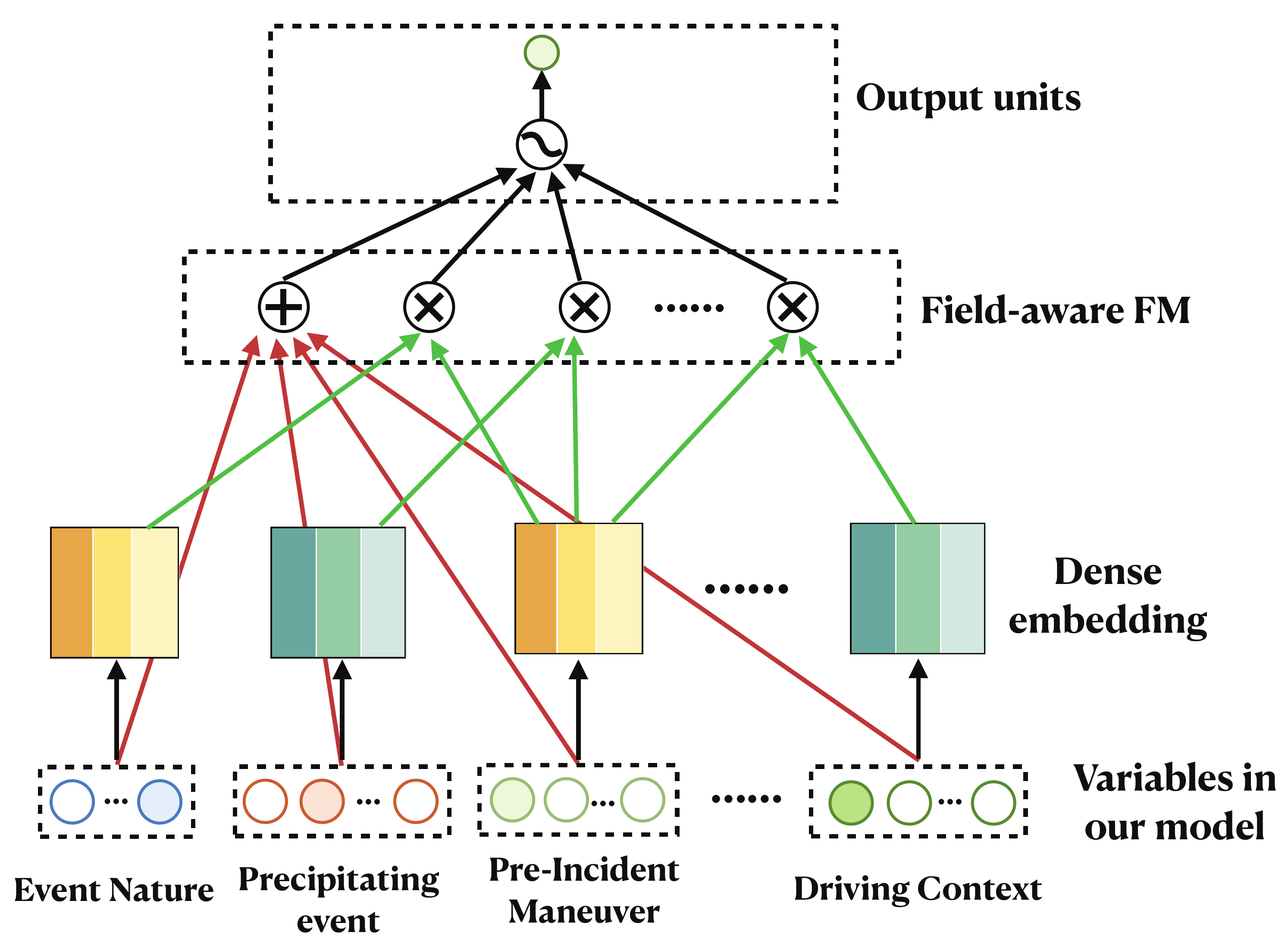}}
\vspace{-10 pt}
\caption{The architecture of the FFMTE model.}
\label{FFM}
\end{figure}

This section introduces the FFMTE model for the traffic event modeling.  As the event variables are correlated with each other, we build our model by considering both the variables and their interactions. The architecture of the FFMTE is shown in Figure \ref{FFM}. 
The core of the model is a factorization machine (FM) \cite{rendle2010factorization}. 
Following \cite{juan2016field}, FFMTE learns both low- and high-order feature interactions. Considering the latent interaction effect of different values for different variables may be different, such as the same evasive maneuvers on icy road or in  extremely dense traffic context, we group all the values for a specific variable into a specific field. 
For each potential value $x_{i}$ for a specific variable in a specific field $F(i)$, a scalar $w_{i}$ is employed to weigh its first-order relevance. Moreover, each value $x_{i}$ has several dense embedding vectors $V_{i}\in \mathbb{R}^{d}$. Depending on the field of other features, one of them $V_{i,F(j)}$ is used to model the second order feature interactions.
All parameters, including $w_{i}$ and $V_{i}$, are trained jointly for the combined prediction model:
\begin{equation}
    f(x) = w_{0}+\sum_{i=1}^{d}w_{i}x_{i} + \sum_{i,j} <V_{i,F(j)},V_{j,F(i)}>x_{i}x_{j}
    \label{ffm_equation}
\end{equation}
where $d$ is the number of variables, and $x_{i}$ is the value of a specific variable in event case $c$. We optimize the FFMTE model using the log-loss function as follows:
\begin{equation}
    \mathcal{L} = \frac{\lambda}{2}||V||_{2} + \sum_{i=1}^{K} log(1+exp(-y_{i}f(x)))
    \label{ffm_equation}
\end{equation}
where $K$ is number of samples, the $\lambda$ is used to perform $L2$ regularization.

For the NDS, the near-crashes are generally several times more than the crash cases (crashes are comparatively rare, as shown in Figure \ref{fig:distribution_data}) \cite{chen2019driving}. To improve the performance of traffic event modeling on these imbalanced datasets, we use the synthetic minority oversampling technique (SMOTE)\footnote{The python implementation    of SMOTE is here: \url{https://imbalanced-learn.org/stable/references/generated/imblearn.over_sampling.SMOTEN.html}} \cite{chawla2002smote} to augment the dataset to make it balance between two categories of data. Training on this augmented dataset using Eq. \ref{ffm_equation}, each traffic event is predicted as crash or near-crash. 

FFMTE shows several benefits for this task: first, it generates the latent embedding vectors $V_{i}$ for the variables in the model, which can be further used in the deductive reasoning process of case retrieval. These vectors facilitate the similarity calculation (e.g., cosine similarity) for ranking the relevant cases in the database (see Section \ref{sec:CBR_Personalized_DA}). 
Second, it models the traffic event much more efficient even when the dataset is quite sparse. FFMTE can train latent vectors $V_{i}$ and $V_{j}$ across multiple data records, not requiring features $i$ and $j$ both appear in the same data record. Therefore, variable interactions in the traffic events, which are never or rarely appeared in the training data, are better modeled by FFMTE \cite{rendle2010factorization}, an advantage contributing to its high performance.

\subsection{Case Base Building}

Building the case base is the primary step to conduct the CBR. Considering the significant value of near-crash event cases to provide the instruction information for future driving, we store all such cases into the overall case base. Ideally, the case base should include all kinds of possibilities of near-crashes in the risk scenarios. However, for the existing collected naturalistic driving data, not all the required cases are presented in the public dataset. 

To overcome this issue, we generate all the possible cases by combining the different potential values for all the variables (i.e., generate all the potential risk scenarios), and use the trained FFMTE model to classify these events. 
Benefiting from the merit of FFMTE model that learns the embedding vectors for all the variables across multiple records, it can be well applied to the never or rarely appeared cases in the training data. Hence we can obtain the set of the classified near-crash cases from all the generated cases, and use them to build the overall case base. Moreover, we store the personal driving history data into a personal case base, based on which the further personalized assistance is provided.

\subsection{CBR for On-board Personalized Driving Assistance\label{sec:CBR_Personalized_DA}}

\begin{figure}[!tbp]
\centerline{\includegraphics[width=0.5\textwidth]{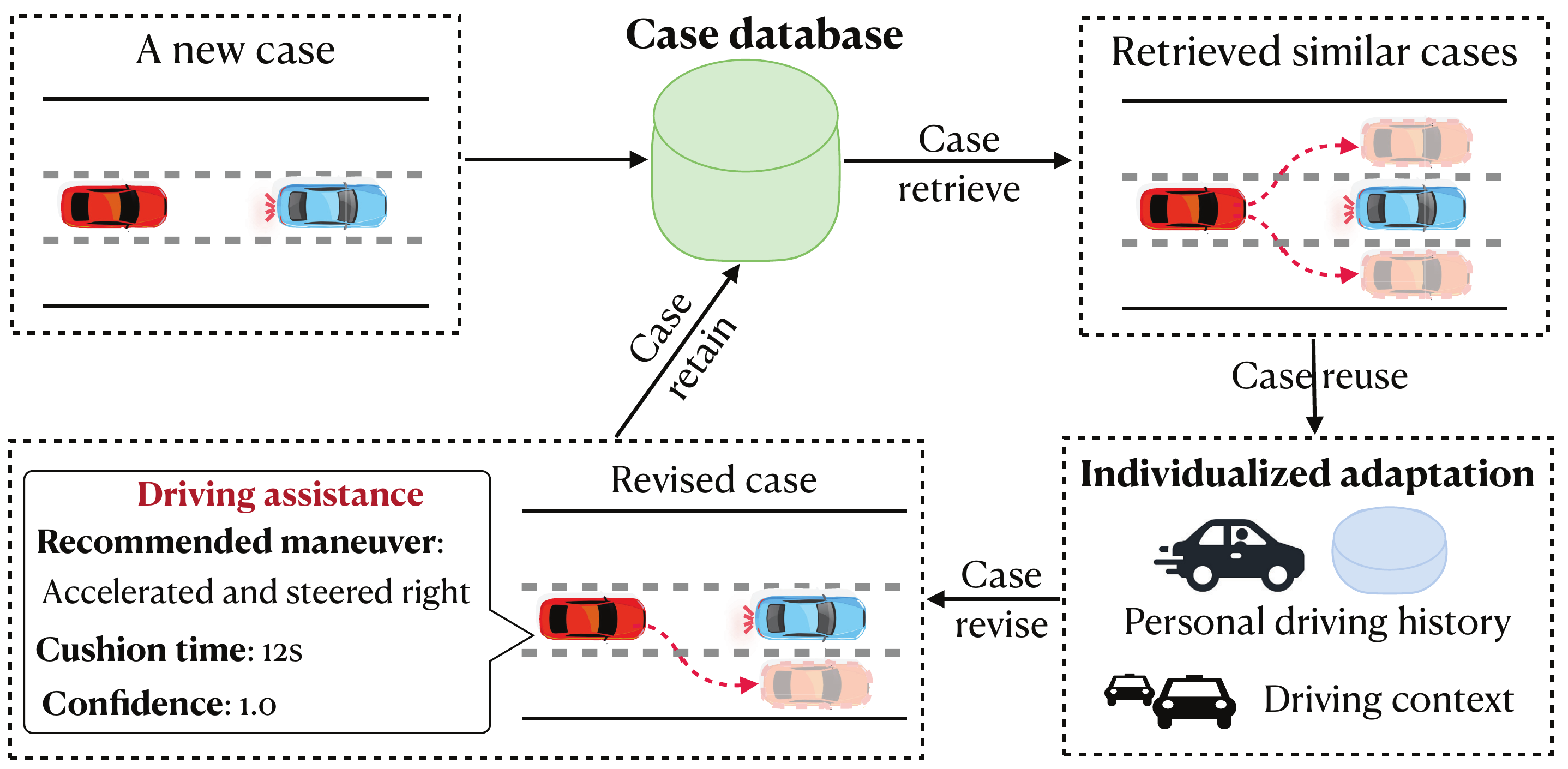}}
\caption{The CBR process for generating personalized driving assistance. Retrieving from the case base using the new case, several similar cases are obtained. Reusing and revising the obtained cases with individualized adaption, the personalized driving assistance can be generated. }
\label{fig:case_based_reasoning}
\end{figure}

CBR is the process of addressing new problems based on prior solutions to similar problems.
For the task in this paper, we rewrite the case representation as $c = \{p,s\}$, where  $p= \{e\_n, p\_e, p\_m, r\_c, d\_c \}$ is the premise of a case and $s=\{d\_r, c\_t\}$ is the solution of the case. 

When a new risk case is encountered, CBR will (1) retrieve the similar existing cases in the case base, and rank the retrieved cases as: $c_{i}>^{+}c_{j}>^{+}...>^{+}c_{k}$, where the partial order $c_{i}>^{+}c_{j}$ suggests that the case $c_{i}$ is more similar to the current case than $c_{j}$ based on the similarity calculation $sim(p, p_{i})>sim(p, p_{j})$; (2) reuse the solutions $s_{i}>^{+}s_{j}>^{+}...>^{+}s_{k}$ by mapping it to the current case; (3) revise the proper solution by adapting it to address the current case; (4) retain the adopted solution $\{p,s\}$ to case base if the adaptation to the current case is successful.
An example of the CBR process for generating personalized driving assistance is shown in Figure \ref{fig:case_based_reasoning}.

\noindent \textbf{Case retrieval:} we define the similarity of two cases by using cosine distance on the dense embedding vectors of the corresponding variables in the premise parts of two cases, as follows:
\begin{equation}
    sim(p_{i},p_{j}) = \frac{1}{N}\sum_{k=1}^{N} cos(V_{i,k},V_{j,k})
\end{equation}
where the $V_{i,k}\in \mathbb{R}^{d}$ is the embedding vector of the $k_{th}$ variable in $p_{i}$, and $N$ is the total numbers of variables in the premise part.
Based on this similarity, the cases in the case base are ranked, and the toppest ranked case shows the most similarity with the current situation. As shown in Figure \ref{fig:case_based_reasoning}, the new case (conflict with a forward decelerating vehicle) have several similar cases (steering right, steering left, etc.) after retrieving from the case base. 

\noindent \textbf{Case reuse:} this step adapts the solutions in the retrieved cases and constructs a set of candidate solutions $\langle\{p,s_{i},score_{i}\},...,\{p,s_{j},score_{j}\}\rangle$ that could be applicable to the current situation, each paired with a confidence level (similarity score).

\noindent \textbf{Case revise:} the candidate solutions are revised according to three empirical criterions: 
\begin{itemize}
    \item the final adopted solution has the highest confidence level;
    \item the solution appears in the personal driving history database;
    \item the solution does not conflict with the current driving context. 
\end{itemize}
The first criterion ensures the adoptability of the solution for the current situation. The second one is for the individualized adaption, it provides the personalized solution that is best fit to the individual driver. Drivers have different preferences and familiarity with different evasive maneuvers, they tend to select the most proficient one in the risk scenarios. Hence we generate the personalized driving assistance by referring their own driving histories, and selecting the most frequent solutions in the candidate sets. The third one is to consider the driving context in the current situation, for example, a driver prefers to steer left in the situation as in Figure \ref{fig:case_based_reasoning}, if the left adjacent lane has another vehicle, the select maneuver of steering left may lead to a crash, hence the final solution should also consider the driving context and do not conflict with the current context.

The personalized driving assistance is then generated for a specific driver. After validating its effectiveness in the realistic scenario, the confirmed case is then retained into the case base for future usage. 

\subsection{Scalability of the Framework}

The proposed framework has good scalability as it is an open evolving framework. With increasingly driving event cases during the naturalistic daily driving accumulated
into the databases, it will evolve with more powerful ability and accuracy to provide the personalized driving assistance. The increasing cases can be used to retrain the traffic event model to improve the model performance constantly. Moreover, the increasingly accumulated personal driving history data also provides more reference cases to instruct the framework to generate more individualized assistance.  
This paper takes the 100-Car NDS dataset as an implementation of the framework to validate its effectiveness in the next section, however, any other naturalistic driving event data can also be embedded into this framework to enrich the case base for scalable extension, we leave this for our future work.  

\section{Experiments}
To validate the efficiency of the framework, we conduct experiments on the 100-Car NDS dataset. Our experiments aim to answer the following questions: (1) how does the FFMTE model perform on event case classification? (2) how well is the CBR results for generating the personalized driving assistance? 

\subsection{Experimental Setup}
 
The original imbalanced dataset is augmented with three methods and formed three balanced datasets: the SMOTE dataset, the CTGAN dataset\footnote{CTGAN for tabular data synthesis: \url{https://sdv.dev/SDV/user_guides/single_table/ctgan.html}}, the random oversampling dataset, each with the same number (760) of crash and near-crash event data. CTGAN uses the generative adversarial networks (GAN) based data synthesizer to learn the data distribution and generate the minority samples \cite{xu2019modeling}.  
We trained the FFMTE model for traffic event modeling using five-fold cross validation on the augmented dataset. 
Grid-search over heuristic choices of hyper-parameters is performed over choices of learning rate: \{0.001, 0.01, 0.05, 0.1, 0.2\},  $\lambda$: \{0.002, 0.02, 0.2\}, training epoch: \{100, 200, 300, 400, 500\}, optimization method: \{sgd, adagrad\}, the dimension $d$ of the embedding vectors: \{3, 4, 5, 6, 7, 8, 9, 10\}. 
After obtaining the best parameter setting for our model (learning rate: 0.05, $\lambda$: 0.002, epoch: 100, adagrad optimization, $d$: 4), we retrain the model on the augmented dataset and test its performance on the original dataset. 

To verify the effectiveness of the proposed FFMTE model, we compared with several baseline models that have been shown to be effective in previous driving event classification researches: the decision tree (DT) \cite{wang2015driving}, support vector machine (SVM) \cite{abdelrahman2022robust}, random forest (RF) \cite{abdelrahman2022robust}, linear regression (LR) \cite{saito2021context}. All the models were implemented in Python, we used xLearn\footnote{xLearn for FFM task: \url{https://xlearn-doc-cn.readthedocs.io/en/latest/#}} as wrapper for the basic FFM; For the other models, the optimal hyperparameters are tuned to report the best results.  
Three criteria are employed to evaluate different facets of the model performance: the accuracy, Area Under Curve (AUC) and the F1. 
The F1 score, a weighted harmonic mean of precision and recall, measures the model performance more comprehensively. 
Moreover, to validate the CBR results, we performed case study to show the generated driving assistance.   

\subsection{Performance on Event Case Classification}

The different models were evaluated by their performances in predicting the event severity on the 100-Car NDS dataset. We trained all the models on the enhanced dataset, and test their performance on the original dataset. 
As show in Table \ref{tab:Performance_comparison}, the proposed FFMTE model achieves the best performance on all the three criteria, with AUC score of 0.981579 and F1 score of 0.981747. The RF obtains the second best, with AUC scored 0.952051 but F1 0.851351. The LR gets the worst on this task, with F1 score only at 0.225, this is probably because the traffic event modeling is not a linear problem, as shown in Figure \ref{cramers_V_correlation}, the valuables are correlated with each other but linear model does not consider such correlations. The high performance of the FFMTE model on this task may attributes to its unique characteristic to conduct efficient factorization on the sparse dataset, it learns the dense embedding vectors crossing multiple data records, hence the event data that is never or rarely appeared in the
training data are better modeled by FFMTE. Moreover, compared with all the baseline models, the FFMTE model can generate the embedding vectors for all the potential values of the considered variables, a by-product which facilitates the case retrieval in the CBR process. 

\begin{table}[!t]
\centering
\caption{Performance comparison of different models.}
 \vspace{-5 pt}
\label{tab:Performance_comparison}
\begin{tabular}{cccc}
\hline
Model             & AUC      & ACC      & F1       \\ \hline
DT     & 0.938661 & 0.973430 & 0.847222 \\
RF    & 0.952051 & 0.973430 & 0.851351 \\
SVM               & 0.908707 & 0.955314 & 0.758170 \\
LR & 0.902545 & 0.925121 & 0.225000 \\
FFMTE               & \textbf{0.981579} & \textbf{0.981579} & \textbf{0.981747} \\ \hline
\end{tabular}
\end{table}

\begin{table}[!t]
\centering
\caption{Performance of the FFMTE model trained on different augmented datasets.}
 \vspace{-5 pt}
\label{tab:FFMTE_different_dataset}
\begin{tabular}{cccc}
\hline
\multicolumn{1}{c}{Dataset} & AUC      & ACC      & F1       \\ \hline
Original dataset   & 0.896401 & 0.981884 & 0.878049 \\
SMOTE dataset     & \textbf{0.981579} & \textbf{0.981579} & \textbf{0.981747} \\
CTGAN dataset      & 0.890687 & 0.890707 & 0.891629 \\
Random oversampling dataset & 0.893270 & 0.893325 & 0.894907 \\ \hline
\end{tabular}
\end{table}

We also compared the different data augmentation techniques on the model performance. Table \ref{tab:FFMTE_different_dataset} shows the performance of the FFMTE model trained on different augmented datasets. The SMOTE method best fits our task, and shows about 11\% improvement compared with the original data. Although the CTGAN method proves some improvement on various tasks, it does not show great lifting on this task, only 2\% on the F1 score. The random oversampling method also shows little improvement.

\subsection{CBR Results for Personalized Driving Assistance}

\begin{table}[!t]
\centering
\caption{CBR results of the three cases in Figure \ref{fig:three_CBR_results}.}
 \vspace{-5 pt}
\label{tab:CBR_3cases_}
\resizebox{\linewidth}{!}{ %
\begin{tabular}{ccc}
\hline
Query Case & \# of General Solutions (thresholded)      & \# of Personalized Solutions           \\ \hline
Case (a)   & 42 & 2  \\
Case (b)     & 56 & 3  \\
Case (c)      & 28 & 2 \\ \hline
\end{tabular}}
\end{table}

Based on the 100-Car NDS dataset, we generated 1,034,880 possible cases by
combining the different potential values for all the variables, and obtained 858,775 near-crash cases after being classified by the high-performance FFMTE model. We used all the 760 near-crash cases in the original dataset to build the personal case database, with the average number of 7.25 event cases collected for a specific driver (1$\sim$52 near-crash cases per driver).

\begin{figure}[!tbp]
\centerline{\includegraphics[width=0.5\textwidth]{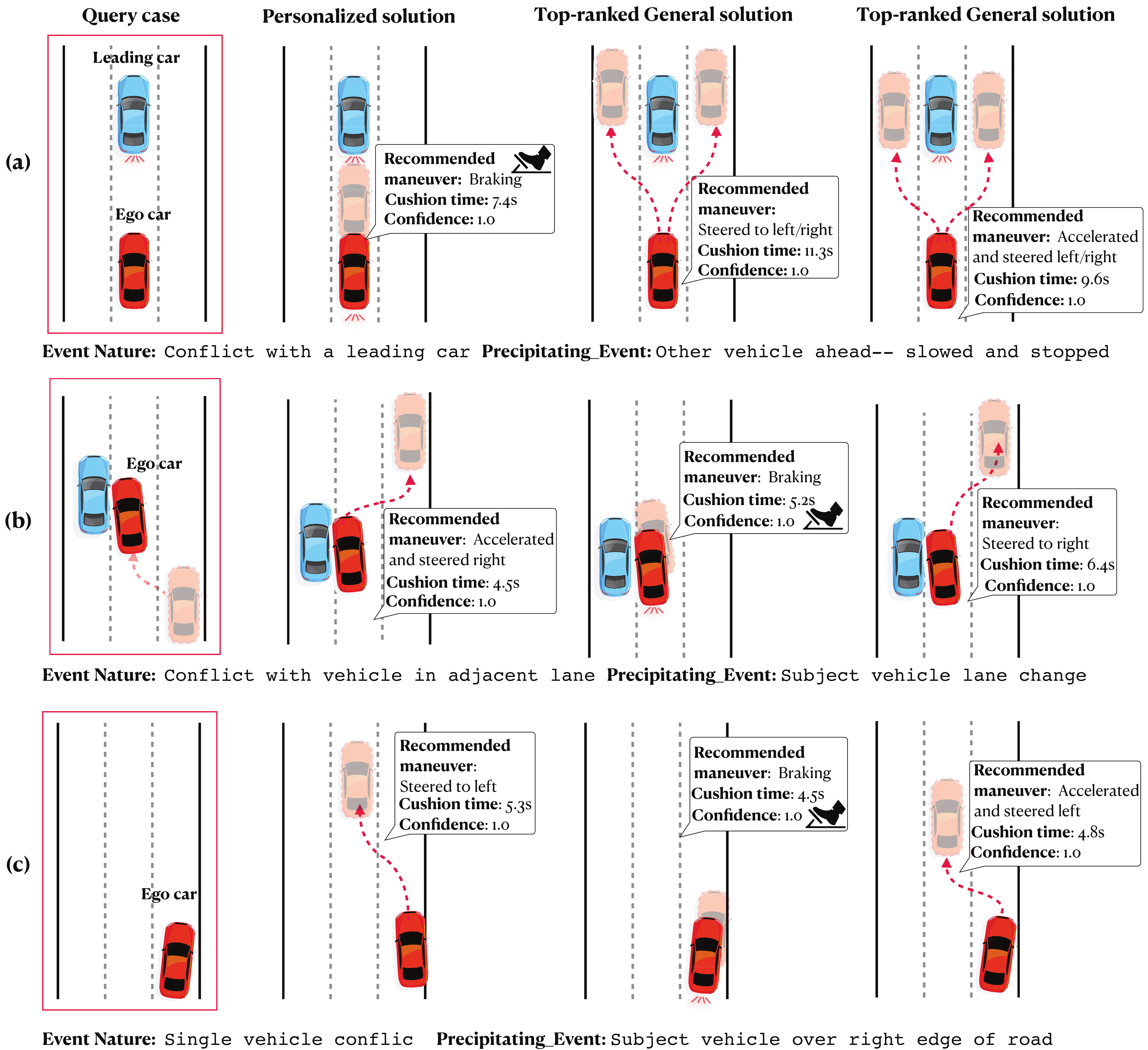}}
 \vspace{-10 pt}
\caption{Three CBR results for generating personalized driving assistance in three different risk scenarios. The three query cases are shown in the first column, the adopted personalized solutions are in the second column, the third and fourth columns are the top-ranked general solutions.} 
\label{fig:three_CBR_results}
\end{figure}

To test the effectiveness of the CBR process for personalized driving assistance on the built case bases, we conducted three case studies using three query cases, as illustrated in Figure \ref{fig:three_CBR_results}. The three query cases were occurring in different risk scenarios: conflict with the leading vehicle (a), conflict with the vehicle in adjacent lane (b), and single vehicle conflict (c). The numbers of thresholded proper general solutions and the personalized solutions are shown in Table \ref{tab:CBR_3cases_}. It shows reasonable retrieved results, the general solutions in the case base provide the feasible driving assistance based on the data from all the drivers, while the personalized solutions retrieved from the personal case database greatly reduce the number of recommended solutions by
fitting the preference of individual drivers. 

Based on the generated personalized solutions, drivers are alerted in proper cushion time to take the proper evasive maneuvers, thus greatly avoiding the potential crashes in different risk scenarios.

\section{Conclusion}
This paper presented an open CBR-based framework for personalized on-board driving assistance in the risk scenarios by leveraging the wealthy of human driving experience from the steady stream of traffic cases (especially the near-crash cases). This framework infers the optimal crash evasive maneuver and the cushion time to hinder
the crash occurrence. To model the traffic events, we proposed the FFMTE model with high performance to classify these events into crash and near-crash cases; we then built the case base using all the near-crash cases as they provide valuable instruction to avoid the different potential risks. A tailored CBR-based method was proposed to retrieve, reuse and revise the similar cases to generate the personalized on-board driving assistance. With increasingly driving event cases accumulated into the databases, this open framework will be evolving and increasingly precise. We took the 100-Car NDS dataset as an example to build and test our framework. The results showed that the proposed FFMTE model achieves the best performance for the event modeling compared with the other baseline models; moreover, the personalized driving assistance in the CBR experiments also showed reasonable retrieval results, providing the drivers with valuable evasive information to avoid the potential crashes.  

For future work, the driving event information, together with the time-series data collected during these events, will be utilized to conduct more holistic traffic event modeling, and to provide more individualized feedback to help drivers avoid potential risks on the road. Ultimately, we expect that the suggested framework will prove effective in real driving to promote safe driving and mitigate future traffic accidents.




\bibliographystyle{unsrt}
\bibliography{mybib}

\end{document}